\documentclass[nonacm]{acmart}

\usepackage{subcaption}
\usepackage{rotating}
\usepackage{multirow}
\usepackage{colortbl}
\usepackage{tabularx}

\newcolumntype{M}[1]{>{\raggedright\arraybackslash}m{#1}}

\copyrightyear{2025}
\acmYear{2025}
\setcopyright{acmlicensed}\acmConference[CHI '25]{CHI Conference on Human Factors in Computing Systems}{April 26-May 1, 2025}{Yokohama, Japan}
\acmBooktitle{CHI Conference on Human Factors in Computing Systems (CHI '25), April 26-May 1, 2025, Yokohama, Japan}
\acmDOI{10.1145/3706598.3713210}
\acmISBN{979-8-4007-1394-1/25/04}

\begin{document}

\title[Towards an Educator-Centered Understanding of Harms from Large Language Models in Education]{``Don't Forget the Teachers'': Towards an Educator-Centered Understanding of Harms from Large Language Models in Education}

\author{Emma Harvey}
\email{evh29@cornell.edu}
\orcid{0000-0001-8453-4963}
\affiliation{%
  \institution{Cornell University}
  \city{Ithaca}
  \state{New York}
  \country{USA}
}
\author{Allison Koenecke}
\email{koenecke@cornell.edu}
\orcid{0000-0002-6233-8256}
\affiliation{%
  \institution{Cornell University}
  \city{Ithaca}
  \state{New York}
  \country{USA}
}
\author{René F. Kizilcec}
\email{kizilcec@cornell.edu}
\orcid{0000-0001-6283-5546}
\affiliation{%
  \institution{Cornell University}
  \city{Ithaca}
  \state{New York}
  \country{USA}
}

\begin{abstract}
Education technologies (edtech) are increasingly incorporating new features built on large language models (LLMs), with the goals of enriching the processes of teaching and learning and ultimately improving learning outcomes. However, the potential downstream impacts of LLM-based edtech remain understudied. Prior attempts to map the risks of LLMs have not been tailored to education specifically, even though it is a unique domain in many respects: from its population (students are often children, who can be especially impacted by technology) to its goals (providing the correct answer may be less important for learners than understanding how to arrive at an answer) to its implications for higher-order skills that generalize across contexts (e.g., critical thinking and collaboration). We conducted semi-structured interviews with six edtech providers representing leaders in the K-12 space, as well as a diverse group of 23 educators with varying levels of experience with LLM-based edtech. Through a thematic analysis, we explored how each group is anticipating, observing, and accounting for potential harms from LLMs in education. We find that, while edtech providers focus primarily on mitigating \textit{technical} harms, i.e., those that can be measured based solely on LLM outputs themselves, educators are more concerned about harms that result from the \textit{broader impacts} of LLMs, i.e., those that require observation of interactions between students, educators, school systems, and edtech to measure. Overall, we (1) develop an education-specific overview of potential harms from LLMs, (2) highlight gaps between conceptions of harm by edtech providers and those by educators, and (3) make recommendations to facilitate the centering of educators in the design and development of edtech tools.\looseness=-1
\end{abstract}

% http://dl.acm.org/ccs.cfm.



\maketitle

\section{Introduction}
Education technology (edtech) companies are increasingly developing products that use large language models (LLMs) to provide personalized tutoring,\footnote{\url{https://www.khanacademy.org/khan-labs}} create instructional materials,\footnote{\url{https://web.diffit.me/}} teach foreign languages,\footnote{\url{https://blog.duolingo.com/duolingo-max}} provide automated writing feedback,\footnote{\url{https://www.grammarly.com/ai}} and more. Many of these tools are already being used in classrooms even though early research on LLMs has revealed myriad risks, including that they can generate content that is toxic, biased, or simply incorrect \cite{bender_dangers_2021, weidinger_taxonomy_2022, lee2024life}. The downstream impacts of LLM use in education remain understudied, and early research has presented conflicting results on whether LLM-based edtech improves or harms student engagement and learning outcomes \cite{nie_gpt_2024, bastani_generative_2024, lo_engagement_2024, pardos_chatgpt_2024}. \looseness=-1

Educational applications of AI are unique in many ways, requiring edtech designers and developers to understand and account for \textit{domain-specific harms} in their products. Much AI research, including responsible AI research, focuses on predicting uncertain outcomes (e.g., disease diagnosis \cite{Hosny2018-ph, obermeyer_dissecting_2019}) or automating complex tasks (e.g., driving \cite{Yurtsever_2020_Survey, ijcai2017p654}). While prediction can certainly be a component of edtech (e.g., Intelligent Tutoring Systems predict at what point a student has mastered a specific knowledge component to move them on to the next one \cite{feng_can_2008}), education itself is not a prediction problem: its aim is to facilitate learning -- something that is neither automatable nor even directly observable, especially for learning objectives related to critical thinking and social skills \cite{MERCER1996359, schuller_benefits_2005, ERIC_survey}. Further, K-12 education involves working with children, which brings with it a new set of privacy and other ethical considerations \cite{Einarsdottir_2007_research}. Taken together, these unique elements of education imply that research into potential harms from LLM-based edtech tools, like prior work on algorithmic fairness \cite{kizilcec_algorithmic_2022, harvey2024towards} and explainable AI \cite{khosravi_explainable_2022}, must be \textit{tailored to the domain of education} in order to have practical value. Finally, LLM-based edtech does not exist in a vacuum -- it is already being deployed in classrooms where educators and students directly engage with it. To understand potential harms from LLM-based edtech, then, it is crucial to draw on HCI research that considers the implications of these \textit{interactions}, following prior work that has explored how students and educators adopt \cite{ibrahim_understanding_2022, kizilcec2024perceived, viberg2024explains}, understand \cite{tan_more_2024, han_teachers_2024}, and use \cite{park_promise_2024, belghith_testing_2024} LLM-based edtech.
\looseness=-1

\subsubsection*{Contributions.} In this context, we explore two key gaps in understanding the potential harms arising from the use of LLMs in edtech. Through a series of semi-structured interviews with edtech providers (N=6) and educators (N=23), we surface an \textbf{education-specific overview of harms} that edtech providers and educators are currently anticipating, observing, or actively working to mitigate. In addition, we highlight \textbf{gaps between conceptions of harm by edtech providers and those by educators} in order to facilitate the \textit{centering of educators} in the design and development of edtech tools. Critically, these gaps come down to a mismatch in the salience of HCI-related factors: while edtech providers focus primarily on harms that can be measured based solely on LLM outputs themselves, educators are more concerned about harms that require observation of interactions between students, educators, school systems, and edtech to measure.
\looseness=-1

Concretely, we identify potential harms arising from the use of LLMs in edtech across three general categories. \textit{Technical} harms include the generation of toxic or biased content, privacy violations, and hallucinations.\footnote{By a \textit{hallucination}, we mean the generation of ``text that is nonsensical, or unfaithful to the provided source input'' \cite{ji_survey_2023}.} Harms arising from \textit{human-LLM interaction} include academic dishonesty. Finally, harms arising from the \textit{broader impact} of LLMs include inhibiting student learning and social development, increasing educator workload while decreasing educator autonomy, and exacerbating systemic inequalities in education. We find that edtech providers focus primarily on mitigating technical harms -- but that these are the same harms that educators report feeling most equipped to address through their teaching practices. On the other hand, educators report high levels of concern about harms resulting from the \textit{broader impacts} of LLMs, most of which edtech providers currently cannot measure or mitigate. In light of this, we identify \textbf{opportunities for edtech providers to better mitigate potential harms through their design practices}. We also identify \textbf{cases where we believe that potential harms from LLM-based edtech cannot be mitigated through design practices alone} and where responsibility thus falls to other actors, including school leaders, regulators, and researchers.\looseness=-1

In \S\ref{s-background}, we provide an overview of the recent shift towards AI-based edtech along with the frameworks that have been proposed to guide its use. We also outline existing domain-agnostic taxonomies of risks from LLMs, and highlight gaps in these taxonomies that prevent them from being directly applicable to educational contexts. Next, in \S\ref{s-method}, we describe our approach to conducting interviews and a thematic analysis. We outline findings from our interviews with edtech providers in \S\ref{s-results-edtech}, and from our interviews with educators in \S\ref{s-results-educators}. Finally, we close by providing recommendations to guide the design and development of educator-centered edtech in \S\ref{s-discussion}. \looseness=-1

\section{Background and Related Work}\label{s-background}
\textit{Educational technology}, or \textit{edtech}, consists of ``technologies specifically designed for educational use as well as general technologies that are widely used in educational settings'' \cite{cardona_artificial_2023}. Edtech need not be based on AI, or even on computing technology. For example, abaci (invented several millennia BCE), electronic calculators (invented in the twentieth century), and WolframAlpha\footnote{\url{https://www.wolframalpha.com/}} (launched in 2009) are all examples of technology that has been used to help students learn math. Nevertheless, recent advances in AI have sparked increased interest in building and using AI-powered edtech to improve learning outcomes and teaching processes, including within the HCI community \citep[e.g.,][]{zhang_mathemyths_2024, cheng_scientific_2024, leong_putting_2024, lee_dapie_2023, lu_readingquizmaker_2023}.\looseness=-1

\subsubsection*{Large Language Models.}\label{s-prior_taxonomy}
Among the most notable of these AI advances are LLMs, sometimes called \textit{foundation models} \cite{bommasani2022opportunities}, which are models trained on massive amounts of text data scraped from the internet to predict the most probable next token (e.g., word, part of a word) in a sequence of text \cite{radford_improving_2018}. By predicting multiple tokens in a sequence, they can create fluent-sounding text, and are increasingly used for natural language understanding and generation tasks. These capabilities bring significant risks as well, as outlined by \citet{bender_dangers_2021} and \citet{weidinger_taxonomy_2022} in two widely cited taxonomies. We draw on both works as a starting point to understand the potential harms that may arise from the use of LLMs in education, and synthesize the harms they identify in Table \ref{t-taxonomy}.\looseness=-1

\begin{table*}
\begin{small}
\centering
\begin{tabular}{ M{0.1\textwidth} M{0.21\textwidth} M{0.31\textwidth} M{0.3\textwidth}}
%{ M{1.5cm} M{3.2cm} M{4.5cm} M{4.5cm}}
\toprule
 \centering{\textbf{Harm Category}} & \centering{\textbf{Required to Measure}} & \centering{\textbf{Domain-Agnostic Harms}\newline\textit{from existing taxonomies}} & \centering{\textbf{Education-Specific Harms}\newline\textit{from our interviews}} 
\tabularnewline
\midrule
\centering{Technical Harms} & 
Outputs of LLM-based systems &
Toxic content, biased content,
privacy violations,
hallucinations & 
Toxic content, biased content,
privacy violations,
hallucinations \\
\midrule
\centering{Human-LLM Interaction Harms} & 
Interactions between LLM-based systems and students and/or educators &
Malicious use (e.g., abusive content, misinformation), HCI harms (from anthropomorphization of LLMs) &  Academic dishonesty \\
\midrule
\centering{Harms From Broader Impacts} & 
Interactions between LLM-based systems and students, educators, and/or school systems &
Research opportunity cost, environmental harms, socioeconomic harms (e.g., labor impacts, reinforcement of inequitable power structures) &  Inhibiting student learning and social development, increasing educator workload, decreasing educator autonomy, exacerbating systemic inequalities in education \\
\bottomrule
\end{tabular}
\end{small}
\caption{Potential harms arising from the use of LLMs. Domain-agnostic harms are synthesized from \citet{bender_dangers_2021} and \citet{weidinger_taxonomy_2022}. Education-specific harms were raised by the edtech providers and educators that we interviewed. }
\label{t-taxonomy}
\Description{A table describing potential harms from LLMs, with the following columns: Harm Category, Required to Measure, Domain-Agnostic Harms, and Education-Specific Context. The first category is technical harms, which require the outputs of LLM-based systems to measure and include the domain-agnostic harms of toxic content, biased content, privacy violations, and hallucinations as well as the education-specific harm of large-scale privacy violations of minors. The second category is human-LLM interaction harms, which require interactions between LLM-based systems and students or educators to measure and include the domain-agnostic harms of malicious use (abusive content, misinformation) and HCI harms (from anthropomorphization of LLMs) as well as the education-specific harm of academic dishonesty. The third category is harms from broader impacts, which require interactions between LLM-based systems and students, educators, and school systems to measure and include the domain-agnostic harms of research opportunity cost, environmental harms, and socioeconomic harms (labor impacts, reinforcement of inequitable power structures) as well as the education-specific harms of inhibiting student learning and social development, increasing educator workload, decreasing educator autonomy, and exacerbating systemic inequalities in education.}
\end{table*}

While these taxonomies provide comprehensive overviews of the potential harms broadly associated with the use of LLMs, they are domain agnostic. Therefore, we set out to place these risks in the context of LLM-based edtech by exploring how edtech providers and educators are anticipating, measuring, and mitigating harms. In doing so, we draw on a rich collection of work related to understanding potential and actual harms resulting from the use of AI in education.\looseness=-1

\subsubsection*{AI in Edtech.} 
A survey of AI in education (AIED) researchers by \citet{holmes_ethics_2022} identified a variety of ethical concerns related to the use of AI in education, including data privacy, quality of education provided by AI tools, teacher and student agency, and equity in AI-based decision-making processes. More recently, in a systematic review of research into the use of LLMs in education, \citet{yan_2024_practical} identified similar concerns as well as inequality in the form of an outsize focus on the English language in existing research. Similarly, \citet{lee2024life} mapped potential sources of bias stemming from each step in the `life cycle' of an LLM. Many of the issues highlighted by prior work have already been observed in practice: AI-based edtech has been shown to discriminate on the basis of race, gender, disability status, and other factors \cite{baker_algorithmic_2022}; impinge on students' privacy and autonomy \cite{diberardino_anti-intentional_2023, commonsense2023}; and provide inaccurate instruction in tutoring settings \cite{wsj_khan}.\looseness=-1

Nevertheless, educators and AIED researchers have good reason to explore the use of AI in education. In the face of pandemic learning loss and the looming expiration of pandemic relief funds \cite{esser, pandemic_recovery}, AI tools are touted as a relatively inexpensive way to meet learners where they are rather than providing the same lessons or assignments to students with different background knowledge or learning needs \cite{cardona_artificial_2023, dai_lin_jin_li_tsai_gasevic_chen_2023, MEYER2024100199}. AI tools also have the potential to make teachers' jobs easier; for example, by providing feedback and support to students outside of teachers' working hours or handling administrative and other non-instruction responsibilities, such as lesson plan development\footnote{See Microsoft Research India's Shiksha copilot \cite{ms_india}.} and grading\footnote{Writable: \url{https://www.writable.com/}} \cite{cardona_artificial_2023}. \looseness=-1

In this uncertain landscape, government agencies \cite{cardona_artificial_2023, noauthor_guidance_2023}, NGOs \cite{noauthor_ethical_2021}, and researchers \cite{KASNECI2023102274, williamson_time_2024} have put forward frameworks intended to guide the responsible development and adoption of AI-based edtech. The US Department of Education (DOE) \cite{cardona_artificial_2023}, for example, has emphasized the importance of centering `humans-in-the-loop'; designing AI tools to adhere to evidence-based pedagogies; and ensuring that AI tools preserve privacy, are explainable, and do not discriminate. In a framework aimed at the procurers of AI-based edtech, the Institute for Ethical AI in Education \cite{noauthor_ethical_2021} identified a similar set of principles, additionally including that AI-based tools do not hinder learners' autonomy and are only deployed to well-informed participants. While helpful, these frameworks are not specific to LLMs and thus do not address in detail several of the risks raised by \citet{bender_dangers_2021} and \citet{weidinger_taxonomy_2022}, such as the potential for LLM-based edtech tools to hallucinate or contribute to academic dishonesty. Other frameworks that specifically consider LLM-based edtech acknowledge the risks of ``unknown unknowns'' \cite{KASNECI2023102274} associated with the technology and call for a deeper examination of ``uncharted ethical issues'' related to access, equity, and human social connection and intellectual development \cite{noauthor_guidance_2023}. Most recently, \citet{williamson_time_2024} have called for a pause on the adoption of LLM-based edtech in schools until policymakers can develop deeper understandings of its risks and until `responsible AI frameworks' are in place for the design and development of future edtech tools. \looseness=-1

In response to these challenges, new guidance from the US DOE has provided a potential path forward for edtech designers and developers seeking to implement these responsible AI frameworks \cite{cardona_designing_2024}. The DOE's report puts forward the ideal of ``designing for education,'' which involves edtech providers and educators engaging in a \textit{co-design process} that uses evidence-based practices to improve teaching and learning. Importantly, the DOE's report highlights the need to \textit{build trust} between edtech providers and educators as a crucial first step in designing for education \cite{cardona_designing_2024}. In our work, we seek to facilitate this trust-building by providing a transparent understanding of how both edtech providers and educators are anticipating, observing, and accounting for potential harms from LLM-based edtech, creating an opportunity for both groups to understand each others' viewpoints and pointing to gaps in current harm mitigation practices. Ultimately, our hope is that this can facilitate the \textit{centering of educators} in the future design and development of edtech tools, and serve as a foundation upon which user-centered and co-design research can build.\looseness=-1
\section{Method}\label{s-method}

\renewcommand{\arraystretch}{1.2}
\begin{table*}
\centering
\begin{small}
\begin{subtable}[t]{0.425\textwidth}
\begin{tabular}{c|c|lc} 
\toprule
\multirow{8}{*}{\begin{tabular}[c]{@{}c@{}}Professional\\Background\end{tabular}} & \multirow{2}{*}{\begin{tabular}[c]{@{}c@{}}Company\\Type\end{tabular}} & For profit & 3 \\
 &  & Non-profit & 3 \\ 
\cline{2-4}
 & \multirow{2}{*}{\begin{tabular}[c]{@{}c@{}}Tool's \\Subject\end{tabular}} & STEM & 3 \\
 &  & Multiple & 3 \\ 
\cline{2-4}
 & \multirow{4}{*}{\begin{tabular}[c]{@{}c@{}}Tool's LLM\\Functionality\end{tabular}} & Live chat & 3 \\
 &  & Live feedback & 2 \\
 &  & Content generation & 3 \\
 &  & Report generation & 2 \\ 
\hline
\multirow{4}{*}{\begin{tabular}[c]{@{}c@{}}Demographic \\Background\end{tabular}} & \multirow{2}{*}{Race} & White (not Hispanic) & 3 \\
 &  & Middle Eastern & 1 \\ 
\cline{2-4}
 & \multirow{2}{*}{Gender} & Man & 2 \\
 &  & Woman & 2 \\ 
\hline
\multirow{3}{*}{\begin{tabular}[c]{@{}c@{}}LLM \\Background\end{tabular}} & \multirow{3}{*}{\begin{tabular}[c]{@{}c@{}}Prior LLM\\Opinion\end{tabular}} & Positive & 4 \\
 &  & Mixed & 2 \\
 &  & Negative & 0 \\
\bottomrule
\end{tabular}
\caption{Edtech providers}
\label{t-interviewees-edtech}
\Description{A table describing the professional, demographic, and LLM backgrounds of the edtech providers we interviewed. 3 participants worked at for-profit companies, and 3 participants worked at non-profit companies. 3 participants had STEM-focused tools, and 3 participants had tools focused on multiple subjects. In terms of LLM functionality, 3 participants were implementing live chat, 3 were implementing live feedback, 3 were implementing content generation, and 3 were implementing report generation. 3 participants identified their race as white, and 1 as Middle Eastern. 2 participants identified their gender identity as a man, and 2 as a woman. 4 participants had positive prior opinions on LLMs, and 2 had mixed opinions.}
\end{subtable}
\hspace{0.1\textwidth}
\begin{subtable}[t]{0.425\textwidth}
\begin{tabular}{c|c|lc} 
\toprule
\multirow{13}{*}{\begin{tabular}[c]{@{}c@{}}Professional\\Background\end{tabular}} & \multirow{5}{*}{Role} & Teacher & 12 \\
 &  & Instructional Support & 6 \\
 &  & Guidance Counselor & 2 \\
 &  & Administrator & 2 \\ 
 &  & Prefer not to say & 1 \\ 
\cline{2-4}
 & \multirow{4}{*}{\begin{tabular}[c]{@{}c@{}}School \\Type\end{tabular}} & Public & 12 \\
 &  & Charter & 6 \\
 &  & Private & 4 \\ 
 &  & Prefer not to say & 1 \\ 
\cline{2-4}
 & \multirow{4}{*}{Subject} & STEM & 10 \\
 &  & Language & 4 \\ 
 &  & Social Studies & 3 \\
 &  & Prefer not to say & 1 \\ 
\hline
\multirow{6}{*}{\begin{tabular}[c]{@{}c@{}}Demographic \\Background\end{tabular}} & \multirow{4}{*}{Race} & White (not Hispanic) & 6 \\
 &  & Black & 3 \\
 &  & Hispanic & 1 \\ 
 &  & Middle Eastern & 1 \\
\cline{2-4}
 & \multirow{2}{*}{Gender} & Man & 6 \\
 &  & Woman & 4 \\ 
\hline
\multirow{7}{*}{\begin{tabular}[c]{@{}c@{}}LLM \\Background\end{tabular}} & \multirow{4}{*}{\begin{tabular}[c]{@{}c@{}}Prior LLM\\Use\end{tabular}} & Significant Use & 3 \\
 &  & Regular Use & 10 \\
 &  & Limited Use & 9 \\
 &  & No Use & 1 \\ 
\cline{2-4}
 & \multirow{3}{*}{\begin{tabular}[c]{@{}c@{}}Prior LLM\\Opinion\end{tabular}} & Positive & 14 \\
 &  & Mixed & 6 \\
 &  & Negative & 3 \\
\bottomrule
\end{tabular}
\caption{Educators}
\label{t-interviewees-educator}
\Description{A table describing the professional, demographic, and LLM backgrounds of the educators we interviewed. 12 participants were teachers, 6 were instructional support staff, 2 were guidance counselors, 2 were administrators, and 1 interviewee preferred not to identify their role. 12 participants worked at public schools, 6 at charter schools, 4 at private schools, and 1 preferred not to identify their school type. 10 participants taught STEM subjects, 4 taught languages, 3 taught social studies, and 1 preferred not to identify their subject. 6 participants identified their race as white, 3 as Black, 1 as Hispanic, and 1 as Middle Eastern. 6 participants identified their gender identity as a man, and 4 as a woman. 3 participants reported significant prior use of LLMs, 10 reported regular prior use, 9 reported limited use, and 1 reported no prior use of LLMs. 14 participants had positive prior opinions on LLMs, 6 had mixed opinions, and 3 had negative opinions.}
\end{subtable}
\end{small}
\caption{Participant backgrounds. All categories above are mutually exclusive and collectively exhaustive, with the following exceptions: all participants were given the option of reporting their race/ethnicity and gender identity and some chose not to; edtech providers reported multiple LLM functionalities within their tools; and we did not record `subject taught' for guidance counselors, administrators, or instructional support staff who did not focus on a specific subject.}
\label{t-interviewees}
\end{table*}

In order to understand how edtech providers and educators are currently anticipating, measuring, and mitigating harms from LLM-based edtech, we conducted 29 semi-structured interviews \cite{Creswell2011} between November 2023 and February 2024. We recruited an initial set of participants via our professional networks and educator listservs. Because STEM teachers were over-represented in our initial recruitment, we recruited additional educators via snowball sampling in order to increase the diversity of participants in terms of their professional roles and academic subjects \cite{morgan2008snowball}. Interviews were between 30 and 60 minutes long and were conducted and recorded via Zoom; participants provided informed consent prior to the start of the recording. We provided all educators interviewed with \$50 compensation for their time.\footnote{We did not provide compensation to edtech providers because, although we recruited educators through listservs and snowball sampling, we recruited edtech providers primarily through a pre-existing research consortium where research participation by both academics and edtech providers is the norm.} Our study protocol was reviewed and deemed exempt by our institution's Institutional Review Board (IRB).\looseness=-1

\subsubsection*{Participants: Edtech Providers.} First, to understand the harms that are actively being anticipated and accounted for in the design and development of edtech tools, we interviewed six individuals employed by distinct edtech providers, identified throughout by IDs T1-T6. These providers represent international leaders in edtech, and each product reports at least 10,000 active student users. The LLM-based functionality that was being developed---or was already deployed---within each edtech tool included: supporting live chat or live feedback,\footnote{\textit{Live feedback} is a short interaction in which an LLM responds to a piece of user-submitted content but does not engage in a longer conversation.} pre-generating content, and creating reports for educators or tool moderators. All participants have research backgrounds and hold roles on their products' leadership or development teams. For-profit, non-profit, general-purpose, and STEM-focused edtech providers are represented in our interview pool. While participants are mostly white and mostly report prior positive opinions on LLMs, we believe that this roughly matches employment trends in the larger tech industry \cite{diversity_tech}. A summary description of the edtech providers we interviewed is provided in Table \ref{t-interviewees-edtech}, and a fuller description of individual edtech providers is provided in Appendix \ref{a-interviewee-characteristics}.\looseness=-1 

Participants are employed at leading edtech providers which are delivering learning materials, assessments, and feedback to millions of students in the US and the UK. However, we are careful not to draw broad generalizations about the universe of edtech providers in this work as our sample comprises only six edtech employees, a pool that is too small to be representative of the large and diverse edtech market. We discuss this in more detail when we consider limitations of our research in \S\ref{s-limitations}. Overall, however, we believe that the practices our interviewees describe are relevant, as they represent the steps that widely used and well-regarded edtech providers take to measure and mitigate harms from LLMs in edtech.\looseness=-1

\subsubsection*{Participants: Educators.} Next, to develop an educator-centered understanding of gaps in existing frameworks for understanding LLM harms in edtech, we interviewed 23 educators, including teachers, instructional support staff, guidance counselors, and school administrators, identified throughout by IDs E1-E23. The educators we spoke with varied in their prior experience with LLMs, including no, limited, regular, and significant\footnote{\textit{Significant use} is defined as use of advanced features, such as creating custom chatbots.} prior use. A summary description of the educators we interviewed is provided in Table \ref{t-interviewees-educator}, and a fuller description of individual educators is provided in Appendix \ref{a-interviewee-characteristics}.\looseness=-1

As with edtech providers, we note that we were not able to recruit a representative sample of educators, and instead prioritized recruiting a \textit{diverse} group in terms of both backgrounds and demographics. As a result, certain backgrounds and perspectives are over-represented in the pool of educators we interviewed. For example, educators who have prior experience with LLMs and educators with a positive opinion of LLMs make up the majority of our sample.\footnote{However, an increasing share of educators within the US broadly have used LLMs: a nationally representative survey conducted by the Center for Democracy \& Technology concurrently with our interviews found that 83\% of educators reported having at least limited personal experience with an LLM, compared with 96\% of our sample \cite{dwyer_up_2024}.} While certain professional and demographic backgrounds are over-represented in our interview pool (white people, men, STEM teachers), we sought to mitigate this by relying on snowball sampling, in which we asked participants to refer us to colleagues of theirs who they believed would have different perspectives from their own \cite{morgan2008snowball}.\looseness=-1

\subsubsection*{Interviews.} 
Participants were first asked to describe their professional roles as well as their personal experience with and opinions (`positive,' `mixed,' or `negative') of LLMs. Participants were then asked an open-ended question about what concerns, if any, they had related to the use of LLMs in edtech (providers were asked about the products that they developed; educators were asked about tools they had used in their classrooms). We used these initial open-ended questions to develop an understanding of the harms that were \textit{most salient} to participants. However, we also specifically wanted to understand whether and to what extent participants were concerned about the set of domain-agnostic harms outlined in Table~\ref{t-taxonomy}. There are two reasons that we chose to ground our interview questions in this set of harms specifically. First, the research that introduces these harms \cite{bender_dangers_2021, weidinger_taxonomy_2022} is widely cited throughout the academic community and represents, to our knowledge, the most common framework for how academics think about potential harms from LLM-based systems. Second, from informal discussions with edtech providers preceeding our interviews, we know that these frameworks also guide how some \textit{edtech providers} themselves think about potential harms from LLMs. Therefore, our goal was to first identify the harms that were most salient to participants (through open-ended questions) and then specifically probe how well widely used domain-agnostic frameworks correspond to education-specific concerns. In order to thoroughly probe opinions on these potential harms, we provided additional prompting if a participant did not proactively raise a potential domain-agnostic harm (e.g., ``are you concerned about toxic or biased content/privacy violations/hallucinations/malicious use/socioeconomic harms?''). Participants were then asked whether and how they had attempted to measure or mitigate any harms that they were concerned about. Once again, the initial question was open-ended and we provided additional prompting if necessary (e.g., for edtech providers, ``have you engaged in red-teaming?'' or for educators, ``have you banned the use of LLMs in your classroom?''). Finally, we asked participants about their future plans for using, and mitigating harm from, LLM-based edtech. Our semi-structured interview guides are available in Appendix \ref{a-interview-guides}.\looseness=-1

The size of our interview pool is commensurate with prior research at CHI on the users and designers of AI and ML tools \cite[e.g.,][]{veale_fairness_2018, chordia_deceptive_2023, scheuerman_products_2024, jin_beyond_2024}, and was large enough for us to conduct interviews until saturation \cite{small_2009_how, hennink_sample_2022}. We defined saturation as occurring when we completed five consecutive interviews in which participants did not proactively raise\footnote{By \textit{proactively raise}, we mean that a participant mentioned a potential harm in response to an open-ended question (e.g., ``what harms are you concerned about?'') and not a direct prompt (e.g., ``are you concerned about hallucinations?'').} a previously unmentioned harm. All of the potential harms from LLMs mentioned by edtech providers were proactively raised by T1; none of T2-T6 mentioned any additional harms. Almost all of the potential harms from LLMs mentioned by educators were proactively raised within the first four educator interviews we conducted, with two exceptions: privacy harms were first proactively raised by E6 and the potential for toxic or biased content was first proactively raised by E18.\footnote{In prior interviews, those harms were discussed, but only when the interviewer specifically prompted educators about them.}\looseness=-1

\subsubsection*{Thematic Analysis.} 
After transcribing the interviews, we used an inductive-deductive coding approach to conduct a thematic analysis \cite{bogdan_qualitative_1998, braun_2006_using}, following similar work in HCI and related fields \citep[e.g.,][]{kawakami_situate_2024, birhane_values_2022}. First, the first author, in discussion with the other authors, developed an initial set of codes based on our pre-identified set of domain-agnostic harms as well as the close-ended questions we asked in our semi-structured interviews. This included participants' professional backgrounds and prior experience with LLMs as well as the harms we specifically prompted all participants to consider if they did not proactively raise them: toxic or biased content, privacy violations, hallucinations, malicious use, and socioeconomic harms. For each harm, we included codes related to (1) whether the harm was proactively raised or was prompted by the interviewer, (2) level of concern about the harm, (3) level of experience with the harm, and (4) planned or implemented strategies for measuring or mitigating the harm. The first author then manually reviewed all transcripts for that initial set of codes. During that review, the first author also noted any additional harms that had been proactively raised by participants. After reviewing all transcripts once, the first author examined all participant-raised harms in order to synthesize related harms; this synthesis was then reviewed and discussed by all authors. The first author then reviewed each transcript a second time, identifying the previously mentioned details for all harms. Finally, the first author examined all identified mitigation strategies across all harms in order to synthesize related mitigation strategies or those that were common across different harms; this synthesis was again reviewed and discussed by all authors.\looseness=-1

We categorized a participant's level of concern as  `lower' or `higher.' In some cases, participants explicitly stated a level of concern (e.g., ``I'm not concerned''); in most cases, we inferred it based on participants' statements about how awareness of that potential harm had impacted or would impact their use of LLMs. For example, in coding statements related to the risk of stereotyping and bias in LLM-generated outputs, we interpreted ``I think I would just work around it'' (E10) to indicate \textit{lower concern} and ``[t]hose are some of the reasons why I'm just not heavy on having ChatGPT in my classroom'' (E7) to indicate \textit{higher concern}.\looseness=-1

\subsubsection*{Positionality.}
We asked participants to describe themselves based on their professional roles as well as their personal experience with and opinions on LLMs. To reflect on our own positionality, we start by doing the same. We are an interdisciplinary team of researchers based at a university in the US. Collectively, we research algorithmic fairness, computational social science, and the learning sciences. None of us have ever been employed as a K-12 educator, but one of us currently holds a research role at an edtech startup; that author also has previously conducted research with both edtech providers and educators. We all identify as having a `mixed' prior opinion on LLMs: we recognize the potential of LLMs to produce benefits in domains including education, but we have also all conducted research about the potential harms of LLMs. Our team is of mixed gender and includes both white and Asian researchers. In terms of our own K-12 education, we have collectively attended public and private schools in the US and internationally.\looseness=-1

We acknowledge that our research has been shaped by this positionality. It informed how we conceptualized this study, how we recruited participants, and how we conducted our thematic analysis. Specifically, our prior experience conducting LLM harm assessments and working with edtech providers informed the set of LLM harm taxonomies we built on, and in turn the questions we asked in our interviews. This is something that we tried to account for by leading with open-ended questions about LLM harms before prompting about specific harms. In addition, we reached our initial pool of participants primarily through the networks afforded to us based on our affiliation with a large research university, although we relied on snowball sampling to broaden the pool. Because our team is based in the US, our interviews and analysis also focus primarily on the US education system.\looseness=-1
\section{How Edtech Providers Account for LLM Harms}\label{s-results-edtech}

The edtech providers we interviewed were aware of and actively working to mitigate technical harms resulting from the use of LLMs in edtech (\S\ref{s-results-technical-harms-edtech}); namely, toxic or biased content, privacy violations, and hallucinations. Only a few edtech providers reported considering harms resulting from human-LLM interactions as they developed tools (\S\ref{s-results-interaction-harms-edtech}), but most reported working to mitigate the potential for LLMs to inhibit student learning (\S\ref{s-results-broader-harms-edtech}). 
\looseness=-1

\subsection{Technical Harms}\label{s-results-technical-harms-edtech}
\subsubsection*{Mitigation Approaches.} All of the edtech providers we interviewed were aware of and attempting to account for the technical LLM harms identified by \citet{bender_dangers_2021} and \citet{weidinger_taxonomy_2022}. Edtech providers reported relying on \textit{human oversight} to measure and mitigate harms, including by internal moderation teams who reviewed content flagged (by toxicity detectors or by users) as being harmful (T1, T5, T6) and by educators who were given access to the chat logs of their students for manual review (T1, T5). Furthermore, multiple providers relied on \textit{technical guardrails}, including LLM-based toxicity detectors (T1, T2, T5, T6), filtering out potential personal information (T2, T4, T5), and using external models or methods like retrieval-augmented generation (RAG) \cite{NEURIPS2020_6b493230} to validate the correctness of LLM outputs (T1, T2). Reliance on pre-existing toxicity detectors like Perspective API\footnote{\url{https://perspectiveapi.com/}} meant that most edtech providers defined toxicity in the same way as those tools: as content that is openly insulting, demeaning, threatening, violent, or sexual; or contains profanity. Edtech providers distinguished \textit{toxic} content from \textit{biased} content (e.g., subtle stereotyping), and generally expressed less certainty about how to mitigate bias as opposed to toxicity, although several edtech providers reported using prompt engineering to explicitly instruct their LLM-based tools to avoid biased responses (T1, T5). \looseness=-1 

\subsubsection*{Open Challenge: Inherently Stochastic Outputs} Importantly, most of the edtech providers we interviewed acknowledged that, due to the inherently stochastic nature of LLMs, their mitigations could not guarantee LLM outputs that were free of toxic or biased content, privacy violations, or hallucinations (T1--T4, T6). Even this small chance of harm was too risky for some participants: ``I only need one student or one teacher to tweet about how [the product] told something very, very inappropriate to a student, and that's it, really'' (T4). Therefore, a key approach to mitigating all of these harms was also to \textit{limit the use of LLMs in some way}; for example, by using LLMs to generate static, human-reviewed content only (T1, T3, T4, T6); by restricting access to LLM agents to only adult users of a tool (T4); and by limiting the context window of live chats to reduce the likelihood of unexpected LLM responses (T1, T3).
\looseness=-1

\subsubsection*{Open Challenge: Dependence on OpenAI}
Finally, several participants discussed challenges related to the ownership of the most powerful LLMs by a small set of companies, most notably OpenAI (T2, T4, T5). The fact that they do not have control over a crucial component of their edtech platforms means that developers must repeatedly test and re-test their prompts and guardrails to ensure that changes to the underlying LLM have not changed their outputs (T2). It also means that tactics like fine-tuning are out of reach for edtech developers who lack the time and resources to re-run their fine-tuning process after each LLM update (T4, T5). \looseness=-1

\subsection{Harms From Human-LLM Interactions}\label{s-results-interaction-harms-edtech}
While the edtech providers we interviewed are working to measure and mitigate technical harms from LLMs, most did not report exploring harms arising from interactions between humans and LLMs. Both of those who did reported using \textit{technical guardrails} to limit the potential for academic dishonesty on their platforms. T1 discussed designing a gate agent to reject adversarial prompts, such as those designed to elicit answers instead of advice from a chatbot. T5 discussed implementing guardrails similar to lockdown browsers to prevent LLMs from being used during assessments.\looseness=-1

\subsection{Harms From the Broader Impacts of LLMs}\label{s-results-broader-harms-edtech}
\subsubsection*{Inhibiting Student Learning} 
Almost every edtech provider we interviewed proactively raised the importance of measuring whether their tool is helpful to student learning (T1--T4, T6). Although there was no consensus on how to best do this, several themes emerged. Collecting and analyzing \textit{user feedback} was the most common approach, with some providers collecting general ratings from all users (T3, T4) and others collecting more detailed reports from small pilot studies (T1, T2). One participant also planned to \textit{compare academic performance} through A/B tests between students who interact with their tool's chatbot and students who interact with older, static versions of their tool as well as pre- and post-performance assessments for students who interact with the chatbot (T1). Another reported measuring helpfulness in a more conceptual way: by performing detailed \textit{risk-benefit analyses} on potential use cases for LLMs before integrating them into the tool (T6). While one participant discussed collecting data from student users in order to verify that helpfulness did not vary across demographic groups (T2), others reported deliberately limiting their access to student data for fear of being seen as `creepy' (T3, T5).\looseness=-1

\subsubsection*{Open Challenge: Adherence to a Pedagogy}
Ensuring that LLM-based edtech tools are contributing to student learning is challenging if LLMs cannot be made to adhere to evidence-based teaching approaches. Multiple participants proactively raised this issue of pedagogy (T1, T4). In particular, when research suggests ``exactly which words to use where and how to present certain material,'' LLM-based tools can provide outputs that deviate from the optimal and recreate ``the average of the internet instructional style'' (T1) -- an issue that the edtech providers we interviewed are still struggling to account for.\looseness=-1

\section{How Educators Account for LLM Harms}\label{s-results-educators}
Like edtech providers, most of the educators we interviewed reported being aware of and working to mitigate technical harms from LLMs, including toxic or biased content, privacy violations, and hallucinations (\S\ref{s-results-technical-harms-educators}). While some educators argued that the potential for LLMs to generate false or biased content rendered them inherently unsafe for use in edtech, most felt confident in their ability to minimize the impact of these harms by \textit{mediating student use of LLMs}. As a result, educators were most concerned about technical harms occurring in cases when LLMs disrupted the teacher-student relationship, as this reduced their ability to personally mitigate harms. Educators felt similarly about harms from human-LLM interaction: although they had high awareness of the potential for LLMs to be used for the purposes of academic dishonesty, they generally reported feeling equipped to identify and address unauthorized use of LLMs by students (\S\ref{s-results-interaction-harms-educators}).\looseness=-1

Educators expressed far less certainty, however, about the broader impacts of LLMs on education (\S\ref{s-results-broader-harms-educators}). This gap between concerns raised by edtech providers and those raised by educators is summarized in Table~\ref{t-gaps}. Specific concerns proactively raised by educators included that LLMs could inhibit students' learning and social development, increase educator workload while decreasing educator autonomy, and exacerbate systemic inequalities in education. Some of these concerns mirror what edtech providers pointed to as their biggest open questions, i.e., issues that they are aware of, but cannot currently measure or mitigate -- and others represent potential harms that edtech providers may not be well-positioned to mitigate, and that may fall to school leaders, regulators, or researchers to address instead.

Again, we note that during interviews, we explicitly prompted all participants to share their level of concern about the potential for LLM-based edtech to produce toxic or biased content, privacy violations, and hallucinations, as well as the potential for misuse\footnote{We used the example of academic dishonesty when prompting about malicious use.} and socioeconomic harms\footnote{We used the example that students from different backgrounds may be differentially-equipped to make use of LLM-based edtech when prompting about socioeconomic harms. However, participants expressed low concern about this particular example and instead proactively raised the other harms that we discuss in \S\ref{s-results-broader-harms-educators}.} if they did not proactively raise these harms. However, we did not prompt participants to discuss any other harms stemming from the broader impact of LLMs: inhibiting student learning and social development, increasing educator workload while decreasing educator autonomy, and exacerbating systemic inequalities in education. All of those harms were instead proactively raised by multiple educators. Therefore, when talking about technical harms and human-LLM interaction harms, we focus on the \textit{level of concern} shared by participants (since all participants were asked about those harms) and when talking about harms from the broader impacts of LLMs we focus on the \textit{salience} to participants (since we only discussed those harms with participants who proactively raised them).\looseness=-1

\begin{table*}
\begin{small}
\centering
\begin{tabular}{ M{0.22\textwidth} M{0.30\textwidth} M{0.2\textwidth} M{0.2\textwidth}}
%{ M{3.5cm} M{5cm} M{2.6cm} M{2.6cm}}
\toprule
 \centering{\textbf{Harm Category}} & \centering{\textbf{Harm}} & \centering{\textbf{Concern Raised by}}\\\centering{\textbf{Edtech Providers}} & \centering{\textbf{Concern Raised by}}\\\centering{\textbf{Educators}}
\tabularnewline
\midrule
& 
Toxic or biased content & 
\centering{\checkmark} & 
\centering{\checkmark} \tabularnewline
\cmidrule{2-4}
\centering{Technical Harms} & 
Privacy violations & 
\centering{\checkmark} & 
\centering{\checkmark} \tabularnewline
\cmidrule{2-4}
& 
Hallucinations & 
\centering{\checkmark} & 
\centering{\checkmark} \tabularnewline
\midrule
\centering{Human-LLM Interaction Harms} & 
Academic dishonesty & 
\centering{\checkmark} & 
\centering{\checkmark} \tabularnewline
\midrule
 & 
Inhibiting student learning & 
\centering{\checkmark} & 
\centering{\checkmark} \tabularnewline
\cmidrule{2-4}
 & 
Inhibiting student social
development & 
 & 
\centering{\checkmark} \tabularnewline
\cmidrule{2-4}
\centering{Harms from Broader Impacts} & 
Increasing educator
workload & 
& 
\centering{\checkmark} \tabularnewline
\cmidrule{2-4}
& 
Decreasing educator
autonomy & 
& 
\centering{\checkmark} \tabularnewline
\cmidrule{2-4}
& 
Exacerbating systemic
inequalities in education & 
& 
\centering{\checkmark} \tabularnewline
\bottomrule
\end{tabular}
\caption{Potential harms from LLM-based edtech that edtech providers and educators proactively raised (i.e., mentioned in response to an open-ended question and not a direct prompt) at least once. While technical harms and human-LLM interaction harms were salient to both edtech providers and educators, harms from broader impacts were more salient to educators.}
\label{t-gaps}
\Description{A table summarizing the concerns that edtech providers and educators independently raised about potential harms from LLMs in education (i.e., without being specifically prompted about the harm). The harms raised independently by both edtech providers and educators were: toxic or biased content, privacy violations, hallucinations, academic dishonesty, and inhibiting student learning. The harms raised independently only by educators were: inhibiting student social development, increasing educator workload, decreasing educator autonomy, and exacerbating systemic inequalities in education.}
\end{small}
\end{table*}

\subsection{Technical Harms}\label{s-results-technical-harms-educators}

\subsubsection*{Toxic Content, Stereotyping, and Bias.}

\begin{quote}
``It was kind of upsetting and confusing, but I saw that happening before I assigned this to [the] class. So we had some debriefing questions at the end, trying to make sense of why that was being the way it was being, the biases that were built in and melded in.'' -- \textit{E1, describing a situation in which an AI tool failed to generate content about non-white people}
\end{quote}

\noindent Educators reported mixed levels of concern about the potential for LLMs used in edtech to generate toxic or biased content (E1, E3-E5, E7, E9, E18, and E19 reported higher levels of concern). Most educators found it deeply unlikely that an LLM used in edtech would generate outright toxic content -- perhaps indicating that edtech providers' mitigation efforts in this regard have been successful. On the other hand, multiple educators had personal experience with biased content, including the generation of ``Eurocentric'' outputs (E1, E13, E18) and negative stereotypes about Black children (E3). While E3 and E18 chose to limit the use of LLMs in their classrooms as a result, E1 and E13, as well as educators who did not report directly experiencing biased outputs (E2, E10), expressed confidence in their abilities to \textit{mediate students' interaction} with these outputs, often by creating `teachable moments' to address bias directly. However, educators' concerns about biased content intensified as they considered situations in which they were not able to serve as intermediaries between students and LLMs, as expressed by E7: ``It's hard enough for me as an individual to stay out of those pitfalls\ldots And when I step into one, I apologize and keep it moving. I can't even fathom having to plan for a computer program making those mistakes\ldots And then I can’t correct that either.''\looseness=-1 

\subsubsection*{Privacy Violations.}
\begin{quote}
    ``Am I concerned about ChatGPT and privacy for children? Yeah. But I'm also concerned about the same thing for them using Facebook\ldots And ultimately that has to be legislated on. I'm not gonna continue worrying about it, because I literally can't do anything about it. But what I can do is, I make every outside vendor that we work with sign a data privacy agreement.'' -- \textit{E13}
\end{quote}

\noindent Educators reported mixed levels of concern about privacy violations resulting from the use of LLMs in edtech (E1, E4-E6, E9, E11, E13, E17, E18, and E20 reported higher levels of concern). Even educators who reported higher levels of concern felt that they could mitigate their concerns by \textit{teaching students how to safely use LLMs without disclosing personal data} (E1, E11, E18). In fact, the most pressing privacy-related concern raised by educators is that students might confide personal details related to abuse or other forms of harm to a chatbot \textit{instead of to an educator} who would be equipped to intervene (E4, E17). In other words, the educators we interviewed are less worried about the potential of LLMs to collect sensitive information about students than they are about the potential that LLMs might prevent educators from accessing that information.\looseness=-1 

\subsubsection*{Hallucinations.}
\begin{quote}
    ``It mitigates a lot of the work that I'll do in the classroom, because then [students] go home, they see this [misconception on ChatGPT], and especially when you're young and a kid, you're like, ‘oh, well, the computer said it.''' -- \textit{E16}
\end{quote}

\noindent The educators we interviewed had more direct experience with hallucinations than with other forms of harm: many educators reported personally noticing a hallucination in LLM outputs (E1-E3, E5, E8, E11, E15, E16, E20, E21). However, personal experience with hallucinations did not seem to translate directly to having a higher level of concern: educators reported mixed concerns overall (E2-E4, E7-E9, E15-E17, E21, and E23 reported higher levels of concern). Hallucination was also the area where educators had the most concrete ideas of how to mitigate harms. Many educators reported that they limited the risk of hallucinations in the first place by only using LLMs to generate static content that they were able to \textit{review before it reached students} (``because we come in assuming that there's going to be some errors, we always double check,'' E18). Educators also reported allowing their students to use LLMs directly, but reducing the impact of harms by \textit{teaching students how to fact-check LLM outputs} (``it's just a new information landscape that we need to train folks on,'' E23). Educators expressed the most concern about situations where students used LLMs without their oversight, i.e., in cases where educators would not be available to correct false information (E7, E16).\looseness=-1

\subsection{Harms From Human-LLM Interactions}\label{s-results-interaction-harms-educators}

\subsubsection*{Academic Dishonesty.}

\begin{quote}
    ``I recently had a teacher [say], `Alice submitted her whole thing, it was all AI generated, I don't know what to do.'\ldots I spoke with her, was like, `your teacher ran this through some AI detection, it came up 98\% AI-generated.' And she's like, `I didn't use AI. I wrote it in Chinese. Google Translate translated it. I put it through Grammarly. My friend corrected it. Then I went back, and --' And then I did a little bit of digging because they wanted to kick her out of the program\ldots and I learned that people are now putting their dissertation from 1994 through AI detection -- it’s like 100\%.'' -- \textit{E5}
\end{quote}

\noindent The educators we interviewed identified one harm arising from human-LLM interactions as particularly relevant in edtech: academic dishonesty (proactively raised as a potential concern by most educators). Although educators had a high \textit{awareness} of the potential for LLMs to be used for academic dishonesty, educators reported mixed levels of \textit{concern} (E2-E6, E10, E15, E17, E19, and E23 reported higher levels of concern). As with the technical harms, this was generally due to the confidence that educators had in their ability to mitigate this risk for their students. For example, educators were more likely to report that they relied on their \textit{intuitions} to identify student use of LLMs (E1, E5, E7, E8, E9, E14, E16, E17, E22, E23) as compared to using \textit{external AI detectors} like Turnitin\footnote{\url{https://www.turnitin.com/products/originality/}} (E2, E3, E8). This was due both to educators' concerns that AI detectors are fallible (E5, see quote above) and to educators' desires for flexibility in how they addressed issues of potential academic dishonesty (``I've developed relationships with these kids\ldots if something comes out on their own accord instead of me blatantly calling the whatever-you-call-it-police, I'll have more time to cultivate that discussion appropriately,'' E4; also E5, E10, E15, E16, E18, E23). Other educators reported \textit{changing their teaching practices} to account for AI, asking ``if a computer can do the task that you assigned, is it the most meaningful task?'' (E21; also E3, E13, E17). Overall, while the educators we interviewed are aware that academic dishonesty is a risk of LLMs, most felt confident in their abilities to mitigate that risk through their teaching practices and interactions with their students.\looseness=-1

\subsection{Harms From the Broader Impacts of LLMs}\label{s-results-broader-harms-educators}

\subsubsection*{Inhibiting Student Learning.}

\begin{quote}
    ``I've noticed that as students become more tech aware, they also tend to lose that critical thinking skill. Because they can just ask for answers.'' -- \textit{E18}
\end{quote}

\noindent The concern that LLMs could inhibit students' learning, especially their critical thinking skills and the development of their own unique `voice,' was proactively raised by educators (E2, E3, E5-E8, E10, E11, E15-E21). While a few educators reported trying to mitigate this issue by \textit{providing opportunities for students to critique LLM outputs} or \textit{consider alternate solutions to those proposed by LLMs during classroom instruction} (E5, E11, E15), most educators either \textit{limited the use of LLMs} in their classroom as a result of these concerns (E3, E7, E8, E10, E16, E17, E18, E19, E20, E21) or expressed that they did not know how to best address them (E2, E6).\looseness=-1 

The edtech providers we interviewed reported trying to address this concern by measuring the `helpfulness' of their tools; however, there is currently no consensus on exactly what `helpfulness' means and how it can be quantified (\S\ref{s-results-broader-harms-edtech}). In addition, like edtech providers, educators raised the issue of pedagogy (E2, E10, E13, E21). Specifically, educators expressed frustration that the LLM-based tools that they have used seem unable to draw vocabulary, teaching processes, and other forms of information from the evidence-based curricula on which they rely in their classrooms. They stressed that this issue can undermine the best practices of educators and confuse students, ultimately lowering overall educational quality even in cases where other harms do not occur.\looseness=-1

\subsubsection*{Inhibiting Student Social Development.}

\begin{quote}
    ``It’s the day and age of the cell phone. So with the AI, I feel like instead of working in groups---cause a lot of kids don't like working in groups---they’ll work with the AI and get the answers. So, human interaction, I think, is going to be cut in half.'' -- \textit{E14}
\end{quote}

\noindent The potential for social harms to students from LLMs was proactively raised by educators (E6, E7, E9, E14, E17-E19). Educators described trying to mitigate this harm by \textit{limiting the use of LLMs} in the classroom, and especially by not assigning LLM-based games or chatbots as part of homework assignments (``Kids now have ways to stay up all hours of the night. And there's nothing we can do. But we don't need to then add to that,'' E7). In general, educators worried that increased reliance on LLMs at school and in daily life would worsen feelings of social isolation in students that educators initially saw emerge as a result of social media and cell phone usage (E14, E19), and that were exacerbated during Covid lockdowns (E6, E7, E9). Educators also expressed worry that overreliance on LLMs could erode relationships and trust between students and teachers (``especially working in juvie, if you don't make that personal connection with your students, they'll shut you out for the rest of your life, because they don't trust you anymore,'' E18, also E17). \looseness=-1 

\subsubsection*{Increasing Educator Workload.}
\begin{quote}
     ``I don't know what's gonna happen to these [AI tools] and I don't know where to invest my own research because I don't have time to look at all of them.'' -- \textit{E5}
\end{quote}

\noindent Educators proactively raised that they felt obligated to spend a significant amount of their professional and personal time researching LLM-based edtech (E1-E6, E8, E9, E12-E15, E18-E21, E23). This included vetting LLM-based tools to determine whether they meet classroom needs (E5, E13, E21), or spending time to learn how to mitigate harms from tools that had been designed and released without their input (E19). Multiple educators also reported attending professional development (PD) sessions (E3, E13, E14, E21, E23) and external courses (E5, E14, E18, E23), or conducting personal or district-sponsored research (E1, E5, E6, E8, E9, E12, E13, E23) in order to better understand the LLM ecosystem. However, the quality of these resources can vary -- many PD sessions are sponsored by edtech providers themselves, and one educator reported receiving inaccurate information when they took an external course (``With the class I’m taking, the professor, the way they talk about AI, it’s like it's foolproof. It’s everything. That's the impression I got. It can't make any mistakes,'' E14).
\looseness=-1 

\subsubsection*{Decreasing Educator Autonomy.}
\begin{quote}
    ``Teachers have to be on board. And teachers’ input about their students is supremely important in creating these tools. Because if a teacher isn’t bought in, they’re not gonna use it with fidelity\ldots So yes, I get principal buy-in and support is important. But don't forget the teachers\ldots The teachers are supremely important in making all of this happen. So listen to them.'' -- \textit{E7}
\end{quote}

\noindent Educators proactively raised the concern that LLM-based edtech tools are not necessarily designed or procured with educator input, or even with concrete educational goals in mind (``I don't know that anybody thought about education when any of these tools came out,'' E1; also E2, E7, E10, E12, E13, E16, E17, E19, E21, E22). In response to the challenges associated with vetting edtech tools, some educators suggested that their time would be better spent providing input into tools directly: ``I would like to give suggestions to what type of responses would be shared, and [help] create something'' (E17). Additionally, multiple educators reported feeling left out of school- or district-wide decisions on LLM procurement in ways that decreased the level of autonomy they felt they had in their classrooms. For example, educators' options are limited when technology vetting occurs at the school, district, or even state level (``Now we're being forced into more bureaucratic apps like Schoolology or ParentSquare and Google Classroom\ldots But now we can't use any of the ones that the kids enjoyed because they weren't vetted,'' E6). At the same time, educators also felt restricted by a lack of clear guidance from their leadership (``The district doesn't know what to do with [generative AI]. So I also have to be careful I don't go too far with it, and then get myself in trouble,'' E8). \looseness=-1 

\subsubsection*{Exacerbating Systemic Inequality.}
\begin{quote}
    ``A lot of these tools that are designed and aimed at teachers -- understandably, they're trying to monetize them. A lot of them are doing the thing where they'll only work at a district level, or they only get a school license. So again, this becomes to me an equity issue\ldots It's hard to feel like it's equitable, or it's gonna be used for public good if it's only available if your district can pony up for it\ldots And I feel like it's always gonna be this income-level gap. Well-off districts are gonna pay for this stuff and struggling districts aren't. And those are the kids that are behind.'' -- \textit{E9}
\end{quote}

\noindent Finally, educators proactively raised the concern that LLM-based edtech could exacerbate systemic inequality by imposing costs that schools and districts are differentially able to bear (E1, E5, E6, E9, E13). Some educators worried about their schools' abilities to shoulder the financial burdens of edtech tools (``it costs a lot of money to get [a tool] vetted by the state,'' E6), while others criticized the system more broadly (``I went to a [professional development session], and it was about how much money edtech platforms make. They actually make more money than curriculum providers, because every educator thinks that they need the curriculum and all these bells and whistles,'' E13).\looseness=-1
\section{Discussion}\label{s-discussion}

Overall, our findings build on previously proposed taxonomies of potential harms from LLMs \cite{bender_dangers_2021, weidinger_taxonomy_2022} by identifying the harms that are most relevant in education: technical harms like toxic or biased content, privacy violations, and hallucinations; interaction harms like academic dishonesty; and harms arising from broader impacts including inhibiting student learning and social development, increasing educator workload while decreasing educator autonomy, and exacerbating systemic inequalities in education. In addition, we highlight gaps between conceptions of harm by edtech providers (who focus primarily on technical harms) and those by educators (who are most concerned about harms resulting from the broader impacts caused by interactions between LLM-based edtech and students, educators, and/or school systems). In doing so, we hope to lay the groundwork for conversations that make the concerns of educators more salient for edtech providers, and at the same time, make the mitigation strategies used by leading edtech designers and developers clear to educators. Our intent is that this work will facilitate the trust-building necessary to ground co-design practices between edtech providers and educators \cite{cardona_artificial_2023}, and lead to the \textit{centering of educators} in the future design and development of edtech tools \cite{kizilcec2024advance}.\looseness=-1

In the remainder of our paper, we discuss our findings in a broader context and point to opportunities for future work. First, we make recommendations to \textit{facilitate the design and development of educator-centered} LLM-based edtech going forward (\S\ref{s-opportunities}). We also reflect on the limitations, ethical considerations, and potential adverse impacts of our work (\S\ref{s-limitations}).\looseness=-1

\begin{table*}
\begin{small}
\centering
\begin{tabular}{ M{0.1\textwidth} M{0.17\textwidth} M{0.27\textwidth} M{0.38\textwidth}}%{ M{1.5cm} M{2.7cm} M{4.1cm} M{5.4cm}}
\toprule
 \centering{\textbf{Harm Category}} & \centering{\textbf{Harm}} & \centering{\textbf{Mitigation Strategies:\\Edtech Providers}} & \centering{\textbf{Mitigation Strategies:\\Educators}}
\tabularnewline
\midrule
\centering{Technical Harms} & 
Toxic or biased content\newline Privacy violations\newline Hallucinations & 
(1) Human oversight,\newline (2) Technical guardrails,\newline (3) Limiting use of LLMs & 
(1) Mediating student interaction with tools: (a) critiquing LLM outputs, (b) training students on safe LLM use, (c) reviewing LLM-generated content before it reaches students;\newline (2) Limiting use of LLMs \\
\midrule
\centering{Human-LLM Interaction Harms} & 
Academic dishonesty & 
Technical guardrails & 
(1) Mediating student interaction with tools: (a) directly addressing suspected academic dishonesty, (b) changing teaching practices to account for AI capabilities;\newline (2) Technical guardrails: (a) AI detectors, (b) lockdown browsers \\
\midrule
 & 
Inhibiting student learning & 
Measuring tool helpfulness: (a) reviewing user feedback, (b) A/B testing users' academic performance, (c) risk-benefit analysis & 
(1) Mediating student interaction with tools:  providing opportunities to critique LLM outputs or consider alternate solutions to those proposed by LLMs;\newline (2) Limiting use of LLMs \\
\cmidrule{2-4}
& 
Inhibiting student social
development & \textit{None surfaced} & 
Limiting use of LLMs \\
\cmidrule{2-4}
\centering{Harms From Broader Impacts} & 
Increasing educator
workload & \textit{None surfaced} & \textit{None surfaced} \\
\cmidrule{2-4}
& 
Decreasing educator
autonomy & \textit{None surfaced} & 
Educating themselves on the LLM ecosystem: (a) attending professional development sessions or external courses, (b) conducting independent research \\
\cmidrule{2-4}
& 
Exacerbating systemic
inequalities in education & \textit{None surfaced} & \textit{None surfaced} \\
  \bottomrule
\end{tabular}
\caption{Mitigation strategies that edtech providers and educators reported practicing to address harms from LLMs in education---and gaps in those strategies.}
\label{t-mitigations}
\Description{A table summarizing the mitigation strategies that edtech providers and educators reported practicing to address harms from LLMs in education. To address toxic or biased content, privacy violations, and hallucinations, edtech providers rely on: 
(1) human oversight, (2) technical guardrails, and (3) limiting their use of LLMs. Educators (1) mediate student interaction with tools by: (a) critiquing LLM outputs, (b) training students on safe LLM use, and (c) reviewing LLM-generated content before it reaches students; and (2) limit their use of LLMs. To address academic dishonesty, edtech providers use technical guardrails. Educators (1) mediate student interaction with tools by: (a) directly addressing suspected academic dishonesty and (b) changing teaching practices to account for AI capabilities; and (2) rely on technical guardrails including : (a) AI detectors and (b) lockdown browsers. To address the harm of inhibiting student learning, edtech providers measure tool helpfulness by: (a) reviewing user feedback, (b) A/B testing users' academic performance, and (c) performing risk-benefit analysis. Educators (1) mediate student interaction with tools by providing opportunities to critique LLM outputs or consider alternate solutions to those proposed by LLMs and limiting their use of LLMs. Edtech providers did not report mitigation strategies for addressing the harm of inhibiting student social development; educators reported limiting their use of LLMs. Neither group reported mitigation strategies for increasing educator workload. Edtech providers did not report mitigation strategies for addressing the harm of decreasing educator autonomy; educators reported educating themselves on the LLM ecosystem by: (a) attending professional development sessions or external courses and (b) conducting independent research. Neither group reported mitigation strategies for exacerbating systemic inequalities in education.}
\end{small}
\end{table*}

\subsection{Recommendations to Facilitate the Design and Development of Educator-Centered Edtech}\label{s-opportunities}

Our interviews surfaced multiple gaps in harm mitigation strategies, outlined in Table \ref{t-mitigations}, that should be addressed by edtech designers and developers, researchers, regulators, and school leaders going forward. We make the following recommendations:\looseness=-1

\begin{enumerate}
    \item \textbf{Edtech providers should design tools in a way that facilitates educator mediation of LLM harms.} Edtech providers currently focus significant energy on mitigating toxic or biased content, privacy violations, hallucinations, academic dishonesty, and the potential for LLMs to inhibit student learning (i.e., lack of helpfulness). At the same time, however, these are the set of harms that educators report feeling able to mitigate by mediating student interaction with tools. By building opportunities for mediation into tools themselves, edtech providers can increase educator autonomy while facilitating the mitigation of a broad list of harms. A promising avenue is co-design practices that allow educators to control the level of oversight that they have over LLM-based edtech \cite[e.g.,][]{de_laet_surveying_2021}.\looseness=-1
    \item  \textbf{Regulators should develop centralized, clear, and independent reviews of LLM-based edtech.} Educators report an increased workload---and a dearth of accurate, unbiased information---related to identifying, vetting, and otherwise learning about LLM-based edtech tools. We echo previous calls \citep[e.g.,][]{williamson_time_2024} for regulators to vet edtech tools. Regulators should leverage existing organizations, such as the What Works Clearinghouse (WWC) established by the US DOE Institute of Education Sciences, to not only vet these tools but also to create searchable repositories of vetted edtech tools.\looseness=-1
    \item \textbf{Researchers and edtech providers should explore how to entrust tool-building to educators themselves.} Throughout this work, we have focused primarily on LLM harms. However, the educators we interviewed were excited about LLM-based edtech in theory, and listed a variety of ways that they, in an ideal world, would use LLMs; for example, generating lesson plans, aligning them to curriculum standards, and adapting them to students' Individualized Education Programs (IEPs).\footnote{\textit{IEPs} are customized learning plans for students with special needs or disabilities.} Currently, educators describe adapting unspecialized tools to suit these needs (``Whether the content in the...plan is what I want it to be or not, it does spit out a structure that I think is really useful,'' E21), with mixed success (``Sometimes wordsmithing what ChatGPT produces ends up being more work than just writing it,'' E8). This current landscape is the continuation of a well-documented trend in edtech in which educational goals are misaligned with the specific capabilities of the AI/ML solutions that seek to address them in practice \cite{liu_reimagining_2023}. However, multiple educators we spoke to described plans to create custom chatbots (through prompt engineering and fine-tuning) that `spoke the language' of their school and their curriculum in a way that off-the-shelf models could not (E12, E20). This is a promising avenue for future research and practice that is already being studied within the HCI community \citep[e.g.,][]{hedderich_piece_2024, f63ccd0b-5bc1-31b0-aa9a-fbd5ab3ba3cc, https://doi.org/10.1111/bjet.12861}.
    \looseness=-1
    \item \textbf{Regulators and school leaders should prioritize educator-centered procurement practices.} These include, for example, actively soliciting educator input in school procurement decisions as well as ensuring that educators are not penalized for choosing \textit{not} to use their schools' LLM-based edtech tools. Procurement policies should follow existing frameworks that require procurers to conduct risk-benefit analysis to explore how LLM-based edtech will improve existing processes without undermining or marginalizing educators \cite[e.g.,][]{noauthor_ethical_2021}.\looseness=-1
    \end{enumerate}

\subsection{Limitations and Ethical Considerations}\label{s-limitations}
\subsubsection*{Limitations} A primary limitation of our work is that we were only able to interview edtech providers and educators based in the US, the UK, and Canada.\footnote{27 interviewees were based in the US, and one each were based in the UK and Canada.} As such, the harms we surface are those relevant to educators from countries that are English-speaking and WEIRD (Western, educated, industrialized, rich and democratic) \cite{Henrich_Heine_Norenzayan_2010}. Well-documented harms of and inequities in LLMs---in particular, that LLMs display cultural biases \cite{tao2024culturalbiasculturalalignment}, perform worse on so-called `low-resource' languages \cite{nicholas_lost_2023, joshi-etal-2020-state}, and that the labor \cite{noema_workers, wsj_workers} and environmental \cite{png_2022_tensions} costs of building and operating LLMs are not equally distributed---were therefore not surfaced by the edtech providers and educators we spoke to. We thus acknowledge that our results are narrowly focused on education in WEIRD, English-speaking countries despite the fact that there is growing scholarship exploring the use of LLMs in edtech globally \cite{henkel2024effective, choi2024llms}, as well as grappling with how to ensure that those efforts do not recreate colonial harms \cite{Shahjahan_decolonizing_2022, ogunremi_decolonizing_2023, bird_decolonising_2020}.\looseness=-1

Additionally, we were not able to recruit a representative sample of interview subjects -- instead, we sought to recruit employees of \textit{widely used and well-regarded} edtech products and educators with a \textit{diverse set of backgrounds and demographics}. As previously noted, this resulted in a relatively small sample size of edtech providers interviewed (six), and we therefore do not attempt to generalize about standard practices across the universe of edtech providers in this work. However, because the edtech providers we interviewed represent leaders in their field, we do believe that the practices they describe are likely to represent emerging best practices -- and at the very least accurately reflect practices that shape widely used edtech products. Further, the sample size of educators we interviewed (23) is commensurate with prior research at CHI, and we were able to conduct interviews on both populations until saturation \cite{hennink_sample_2022, small_2009_how}.\looseness=-1

\subsubsection*{Ethical Considerations.}
In conducting this work, we faced a classic tension inherent to participatory AI research \cite{feffer_preference_2023, birhane_power_2022, sloane_participation_2022}: our goal with this work was to facilitate the centering of educators in the future development of edtech tools, but our method for doing so (Zoom interviews) placed demands on educators' (already limited) time. To mitigate this, we provided competitive compensation (\$50 per educator, corresponding to an hourly rate of between \$50 and \$100 depending on interview length).\looseness=-1

\subsubsection*{Adverse Impact.}
Our interviewees spoke to us under the condition of anonymity: edtech providers shared potentially sensitive product details and processes with us, and educators shared critical thoughts on their employers and working environments. As such, a major potential adverse impact of our work is the risk that any of our interviewees may be identified. To avoid this, we have taken the following steps: (1) anonymizing all quotes, (2) providing characteristics of our interviewees at only a low level of granularity, (3) storing data securely in accordance with our IRB, and (4) deleting the original meeting recordings after transcribing them. Other than this, we do not believe that any of our findings are likely to be co-opted or used in an adversarial way. \looseness=-1

\section{Conclusion}
Through a series of interviews with edtech providers (N=6) and educators (N=23), we surfaced an \textbf{education-specific overview of LLM harms} that edtech providers and educators are currently anticipating, observing, or actively working to mitigate. These include: technical harms (toxic or biased content, privacy violations, hallucinations), interaction harms (academic dishonesty), and harms from the broader impact of LLMs (inhibiting student learning and social development, increasing educator workload, decreasing educator autonomy, and exacerbating systemic inequalities in education). We find that edtech providers focus almost exclusively on mitigating \textit{technical} harms, which are measurable based solely on the outputs of LLM-based systems -- but that these are the same harms that educators report feeling most equipped to mediate through their teaching practices. On the other hand, educators report high levels of concern about harms resulting from the \textit{broader impacts} of LLMs -- harms that require observing interactions between LLM-based systems and students, educators, and/or school systems to measure. Overall, we provide \textbf{an education-specific overview of potential harms from LLMs}, building on widely used domain-agnostic taxonomies. In addition, we identify \textbf{gaps between conceptions of harm by edtech providers and those by educators}. Finally, we make \textbf{recommendations for edtech designers and developers, researchers, regulators, and school leaders} to bridge those gaps and contribute to the design of \textit{educator-centered} edtech. \looseness=-1

\begin{acks}
We thank all study participants and anonymous reviewers. This work is supported by funding from the Schmidt Futures Foundation as part of the Learning Engineering Virtual Institute (LEVI), and by an award from the National Science Foundation (2237593). 
\end{acks}

\bibliographystyle{ACM-Reference-Format}
\bibliography{references}

\appendix

\section{Semi-Structured Interview Guides}\label{a-interview-guides}
\textit{All participants were first provided with a brief introduction to the study, offered a chance to ask questions, and asked for their informed consent to participate and be recorded:}
\newline\newline\noindent Thank you for taking the time to join this Zoom! [Interviewer introduces themselves.] I am asking you to participate in a research study intended to help us understand how educators and edtech providers are conceptualizing and measuring the potential risks posed by the use of large language models in education, with the goal of creating an educator-centered framework to guide the development of future edtech tools. If you consent to participate in this study, you will be asked to participate in a 30-to-45-minute interview about how you conceive of and measure potential risks posed by the use of large language models in education. We will also create an audio-video recording of the Zoom meeting for the purposes of creating a de-identified transcript of our conversation. Following the transcription, we will destroy the Zoom media recording. The de-identified transcript will be stored securely. Participation in this study is completely voluntary. We do not anticipate any risks associated with participation, and there are no benefits other than contributing to our understanding of this topic. Do you have any questions about this study that I can answer? Do you consent to participate in this research? 
\newline\newline\noindent \textit{Next, we asked a separate set of questions of edtech providers and educators. Because we followed a semi-structured interview process, not all interviewees were asked all questions, the question order varied across interviews, and some interviewees were asked follow-ups not listed here in response to statements that they made.}

\subsection{Edtech Providers}
Background Information: 
\begin{itemize}
    \item Your role on the product development team
    \item Describe the LLM-based tool or product
    \begin{itemize}
        \item Purpose
        \item Intended student population
        \item For-profit or nonprofit 
        \item The backend technology: proprietary, an existing LLM with fine-tuning, an existing LLM with a content moderation wrapper, other? 
    \end{itemize}
\end{itemize}
LLM Background: 
\begin{itemize}
    \item Your additional experience with LLMs
    \begin{itemize}
        \item Do you use them personally?
        \item Which ones? What for? 
    \end{itemize} 
    \item Your attitude towards LLMs 
    \begin{itemize}
        \item Positive? Negative? Mixed?
        \item Has this changed over time?
    \end{itemize}  
\end{itemize}
Risk Identification:
\begin{itemize}
    \item While developing your tool, what potential risks did you identify related to its use? 
    \item If any harm is not mentioned, prompt: 
    \begin{itemize}
        \item Malicious use: plagiarism / cheating
        \item Misinformation: LLMs supplying incorrect information to students
        \item Discrimination, hate speech, and exclusion: LLMs producing toxic content, reproducing harmful stereotypes, or treating different students differently
        \item Information hazards: concerns related to privacy or students sharing sensitive information
        \item Socioeconomic harms: students of different backgrounds being differentially-equipped to make use of the tool    \end{itemize}
    \item If tool has been deployed, are there any additional risks that you identified post-deployment? 
\end{itemize}

Risk Measurement and Mitigation: 
\begin{itemize}
    \item How have you tried to measure or mitigate these risks? 
    \item If interviewee is unsure, prompt: 
    \begin{itemize}
        \item Red-teaming?
        \item User reporting functionality? 
    \end{itemize}
    \item Is there anything that you tried that didn’t work?
    \item Is there anything you plan to do but haven’t tried yet? 
    \item Is there anything that you don’t know how to measure? 
\end{itemize}

Closing:
\begin{itemize}
    \item Is there anything else you’d like to share with us?
\end{itemize}

\subsection{Educators}
Background Information: 
\begin{itemize}
    \item Your role as an educator
    \begin{itemize}
        \item Subject
        \item Grade level
        \item Location
        \item School type (public, private, charter)
    \end{itemize}
\end{itemize}
LLM Background: 
\begin{itemize}
    \item Your prior experience with LLMs
    \begin{itemize}
        \item Have you read about them in the news?
        \item Have you used them in your classroom?
        \item Do you use them personally?
        \item Which ones? What for? 
    \end{itemize} 
    \item Your attitude towards LLMs 
    \begin{itemize}
        \item Positive? Negative? Mixed?
        \item Has this changed over time?
    \end{itemize}  
\end{itemize}
Risk Identification:
\begin{itemize}
    \item What concerns do you have around LLM use in your classroom? 
    \item If any harm is not mentioned, prompt: 
    \begin{itemize}
        \item Malicious use: plagiarism / cheating
        \item Misinformation: LLMs supplying incorrect information to students
        \item Discrimination, hate speech, and exclusion: LLMs producing toxic content, reproducing harmful stereotypes, or treating different students differently
        \item Information hazards: concerns related to privacy or students sharing sensitive information
        \item Socioeconomic harms: students of different backgrounds being differentially-equipped to make use of the tool    \end{itemize}
    \item If interviewee is unsure about potential LLM use cases, prompt to think about harms associated with the following use cases specifically:
    \begin{itemize}
        \item Chatbot-style tutors
        \item Automated lesson plan creators
        \item Training and support tools for teachers
    \end{itemize}
\end{itemize}

Risk Measurement and Mitigation: 
\begin{itemize}
    \item How have you tried to measure or mitigate these risks? 
    \item If interviewee is unsure, prompt: 
    \begin{itemize}
        \item Banning LLM use?
        \item Asking students to report toxic content? 
        \item Anti-plagiarism software? 
    \end{itemize}
    \item Is there anything that you tried that didn’t work?
    \item Is there anything you plan to do but haven’t tried yet? 
    \item Is there anything that you don’t know how to measure? 
\end{itemize}

Future Plans with LLMs: 
\begin{itemize}
    \item In an ideal world where all of your concerns related to LLMs were mitigated, what aspects of your job, if any, would you ideally delegate to an LLM?
    \item If interviewee is unsure, prompt: 
    \begin{itemize}
        \item Administrative tasks
        \item Non-student-facing tasks
        \item Tutoring or automated feedback
    \end{itemize}
    \item How do you plan to use LLMs over the next year (reduce, continue, or expand current use)?
\end{itemize}

Closing:
\begin{itemize}
    \item Is there anything else you’d like to share with us?
\end{itemize}

\section{Interviewee Characteristics}\label{a-interviewee-characteristics}
\begin{table*}[h]
\centering
\begin{tabular}{c|c!{\color{lightgray}\vrule}c|c!{\color{lightgray}\vrule}c|c!{\color{lightgray}\vrule}c!{\color{lightgray}\vrule}c!{\color{lightgray}\vrule}c|c!{\color{lightgray}\vrule}c|c!{\color{lightgray}\vrule}c|c!{\color{lightgray}\vrule}c!{\color{lightgray}\vrule}c|}

\multicolumn{1}{c}{}
& \multicolumn{2}{c}{\textbf{Type}} &
\multicolumn{2}{c}{\textbf{Subject}} &
\multicolumn{4}{c}{\textbf{LLM Use}} &
\multicolumn{2}{c}{\textbf{Race}} &
\multicolumn{2}{c}{\textbf{Gender}} &
\multicolumn{3}{c}{\textbf{Opinion}} \\
\arrayrulecolor{black}\cline{2-16}
& \begin{sideways}\textit{Non-profit}\end{sideways} & \begin{sideways}\textit{For profit}\end{sideways} &
 \begin{sideways}\textit{STEM}\end{sideways} &
 \begin{sideways}\textit{Multiple}\end{sideways} &
 \begin{sideways}\textit{Live Chat}\end{sideways} & \begin{sideways}\textit{Live Feedback}\end{sideways} & \begin{sideways}\textit{Pre-Generated Content\textcolor{white}{..}}\end{sideways} & \begin{sideways}\textit{Report Generation~}\end{sideways} & \begin{sideways}\textit{White}\end{sideways} & \begin{sideways}\textit{Middle Eastern}\end{sideways} & \begin{sideways}\textit{Man}\end{sideways} & \begin{sideways}\textit{Woman}\end{sideways} & \begin{sideways}\textit{Positive}\end{sideways} & \begin{sideways}\textit{Mixed}\end{sideways} & \begin{sideways}\textit{Negative}\end{sideways} \\
\arrayrulecolor{black}\hline
\textit{T1} & & \checkmark &&\checkmark & \checkmark & & \checkmark & & & \checkmark & \checkmark & & \checkmark & & \\
\arrayrulecolor{lightgray}\hline
\textit{T2} & & \checkmark & \checkmark && \checkmark && & & & & & & & \checkmark & \\
\arrayrulecolor{lightgray}\hline
\textit{T3} & \checkmark & & \checkmark && & \checkmark & \checkmark & & & & & & & \checkmark & \\
\arrayrulecolor{lightgray}\hline
\textit{T4} && \checkmark & \checkmark && && \checkmark & \checkmark & \checkmark && \checkmark && \checkmark && \\
\arrayrulecolor{lightgray}\hline
\textit{T5} & \checkmark && & \checkmark & \checkmark &&&& \checkmark &&& \checkmark   & \checkmark   & & \\
\arrayrulecolor{lightgray}\hline
\textit{T6} & \checkmark && & \checkmark & & \checkmark & & \checkmark & \checkmark & & & \checkmark & \checkmark & & \\
\arrayrulecolor{black}\hline
\textbf{Totals} & 3 & 3 & 3 & 3 & 3 & 2 & 3 & 2 & 3 & 1  & 2 & 2  & 4  & 2  & 0 \\ 
\arrayrulecolor{black}\hline
\end{tabular}
\caption{Characteristics of the edtech providers that we interviewed.}
\label{t-interviewees-edtech-individual}
\Description{A table describing the individual professional and demographic backgrounds and prior opinions on LLMs of the edtech providers we interviewed. T1 is a Middle Eastern man who has a positive opinion on LLMs and is employed at a for-profit company whose edtech product focuses on multiple subjects and is using LLMs for live chat and pre-generated content. T2 has a mixed opinion on LLMs and is employed at a for-profit company whose edtech product focuses on STEM and is using LLMs for live chat. T3 has a mixed opinion on LLMs and is employed at a non-profit company whose edtech product focuses on STEM and is using LLMs for live feedback and pre-generated content. T4 is a white man who has a positive opinion on LLMs and is employed at a for-profit company whose edtech product focuses on STEM and is using LLMs for pre-generated content and report generation. T5 is a white woman who has a positive opinion on LLMs and is employed at a non-profit company whose edtech product focuses on multiple subjects and is using LLMs for live chat. T6 is a white woman who has a positive opinion on LLMs and is employed at a non-profit company whose edtech product focuses on multiple subjects and is using LLMs for live feedback and report generation.}
\end{table*}
\begin{table*}[t]
\centering
\begin{tabular}{c|c!{\color{lightgray}\vrule}c!{\color{lightgray}\vrule}c!{\color{lightgray}\vrule}c!{\color{lightgray}\vrule}c|c!{\color{lightgray}\vrule}c!{\color{lightgray}\vrule}c!{\color{lightgray}\vrule}c|c!{\color{lightgray}\vrule}c!{\color{lightgray}\vrule}c!{\color{lightgray}\vrule}c|}

\multicolumn{1}{c}{}
& \multicolumn{5}{c}{\textbf{Role}} &
\multicolumn{4}{c}{\textbf{School Type}} &
\multicolumn{4}{c}{\textbf{Subject}} \\

\arrayrulecolor{black}\cline{2-14}
& \begin{sideways}\textit{Teacher}\end{sideways} & \begin{sideways}\textit{Instructional Support\textcolor{white}{..}}\end{sideways} & \begin{sideways}\textit{Guidance Counselor~}\end{sideways} & \begin{sideways}\textit{Administrator}\end{sideways} &
 \begin{sideways}\textit{Prefer not to say}\end{sideways} &
 \begin{sideways}\textit{Public}\end{sideways} & \begin{sideways}\textit{Private}\end{sideways} & \begin{sideways}\textit{Charter}\end{sideways} & 
 \begin{sideways}\textit{Prefer not to say}\end{sideways} &
 \begin{sideways}\textit{Language}\end{sideways} & \begin{sideways}\textit{Social Studies}\end{sideways} & \begin{sideways}\textit{STEM}\end{sideways} & 
 \begin{sideways}\textit{Prefer not to say}\end{sideways} \\
\arrayrulecolor{black}\hline
\textit{E1} & 
\checkmark &&&&&
& \checkmark &&& 
\checkmark &&& \\
\arrayrulecolor{lightgray}\hline
\textit{E2} & 
&&&& \checkmark &
&&& \checkmark & 
&&& \checkmark \\
\arrayrulecolor{lightgray}\hline
\textit{E3} & 
\checkmark &&&&&
& \checkmark &&& 
\checkmark &&& \\
\arrayrulecolor{lightgray}\hline
\textit{E4} & 
&& \checkmark &&&
& \checkmark &&& 
&&& \\
\arrayrulecolor{lightgray}\hline
\textit{E5} & 
\checkmark &&&&&
& \checkmark &&& 
\checkmark &&& \\
\arrayrulecolor{lightgray}\hline
\textit{E6} & 
&& \checkmark &&&
\checkmark &&&& 
&&& \\
\arrayrulecolor{lightgray}\hline
\textit{E7} & 
\checkmark &&&&&
&& \checkmark && 
&& \checkmark & \\
\arrayrulecolor{lightgray}\hline
\textit{E8} & 
\checkmark &&&&&
\checkmark &&&& 
& \checkmark && \\
\arrayrulecolor{lightgray}\hline
\textit{E9} & 
& \checkmark &&&&
\checkmark &&&& 
\checkmark &&& \\
\arrayrulecolor{lightgray}\hline
\textit{E10} & 
\checkmark &&&&&
\checkmark &&&& 
&& \checkmark & \\
\arrayrulecolor{lightgray}\hline
\textit{E11} & 
\checkmark &&&&&
&& \checkmark && 
&& \checkmark & \\
\arrayrulecolor{lightgray}\hline
\textit{E12} & 
&&& \checkmark &&
&& \checkmark && 
&&& \\
\arrayrulecolor{lightgray}\hline
\textit{E13} & 
&&& \checkmark &&
&& \checkmark && 
&&& \\
\arrayrulecolor{lightgray}\hline
\textit{E14} & 
\checkmark &&&&&
\checkmark &&&& 
& \checkmark && \\
\arrayrulecolor{lightgray}\hline
\textit{E15} & 
\checkmark &&&&&
&& \checkmark && 
& \checkmark && \\
\arrayrulecolor{lightgray}\hline
\textit{E16} & 
\checkmark &&&&&
&& \checkmark && 
&& \checkmark & \\
\arrayrulecolor{lightgray}\hline
\textit{E17} & 
& \checkmark &&&&
\checkmark &&&& 
&& \checkmark & \\
\arrayrulecolor{lightgray}\hline
\textit{E18} & 
& \checkmark &&&&
\checkmark &&&& 
&& \checkmark & \\
\arrayrulecolor{lightgray}\hline
\textit{E19} & 
\checkmark &&&&&
\checkmark &&&& 
&& \checkmark & \\
\arrayrulecolor{lightgray}\hline
\textit{E20} & 
& \checkmark &&&&
\checkmark &&&& 
&& \checkmark & \\
\arrayrulecolor{lightgray}\hline
\textit{E21} & 
& \checkmark &&&&
\checkmark &&&& 
&& \checkmark & \\
\arrayrulecolor{lightgray}\hline
\textit{E22} & 
\checkmark &&&&&
\checkmark &&&& 
&& \checkmark & \\
\arrayrulecolor{lightgray}\hline
\textit{E23} & 
& \checkmark &&&&
\checkmark &&&& 
&&& \\
\arrayrulecolor{black}\hline
\textbf{Totals} & 12 & 6 & 2 & 2 & 1 & 12 & 4 & 6 & 1 & 4 & 3 & 10 & 1 \\ 
\arrayrulecolor{black}\hline
\end{tabular}
\caption{Professional backgrounds of the educators that we interviewed.}
\label{t-interviewees-educator-individual}
\Description{A table describing the individual professional backgrounds of the educators we interviewed. E1 is a language teacher at a private school. E2 chose not to provide details on their professional background. E3 is a language teacher at a private school. E4 is a guidance counselor at a private school. E5 is a language teacher at a private school. E6 is a guidance counselor at a public school. E7 is a STEM teacher at a charter school. E8 is a social studies teacher at a public school. E9 is a language instructional support staff member at a public school. E10 is a STEM teacher at a public school. E11 is a STEM teacher at a charter school. E12 is an administrator at a charter school. E13 is an administrator at a charter school. E14 is social studies teacher at a public school. E15 is a social studies teacher at a charter school. E16 is a STEM teacher at a charter school. E17 is a STEM instructional support staff member at a public school. E18 is a STEM instructional support staff member at a public school. E19 is a STEM teacher at a public school. E20 is a STEM instructional support staff member at a public school. E21 is a STEM instructional support staff member at a public school. E22 is a STEM teacher at a public school.  E23 is an instructional support staff member at a public school.}
\end{table*}

\begin{table*}[t]
\centering
\begin{tabular}{c|
c!{\color{lightgray}\vrule}
c!{\color{lightgray}\vrule}
c!{\color{lightgray}\vrule}
c|
c!{\color{lightgray}\vrule}
c|
c!{\color{lightgray}\vrule}
c!{\color{lightgray}\vrule}
c!{\color{lightgray}\vrule}
c|
c!{\color{lightgray}\vrule}
c!{\color{lightgray}\vrule}
c|}

\multicolumn{1}{c}{}
& \multicolumn{4}{c}{\textbf{Race}} &
\multicolumn{2}{c}{\textbf{Gender}} &
\multicolumn{4}{c}{\textbf{LLM Use}}&
\multicolumn{3}{c}{\textbf{Opinion}} \\

\arrayrulecolor{black}\cline{2-14}
& \begin{sideways}\textit{White (not Hispanic)\textcolor{white}{..}}\end{sideways} & \begin{sideways}\textit{Black}\end{sideways} & \begin{sideways}\textit{Middle Eastern}\end{sideways} & \begin{sideways}\textit{Hispanic}\end{sideways} & \begin{sideways}\textit{Man}\end{sideways} & \begin{sideways}\textit{Woman}\end{sideways} & \begin{sideways}\textit{Significant}\end{sideways} & \begin{sideways}\textit{Regular}\end{sideways} & \begin{sideways}\textit{Limited}\end{sideways} & \begin{sideways}\textit{None}\end{sideways} & \begin{sideways}\textit{Positive}\end{sideways} & \begin{sideways}\textit{Mixed}\end{sideways} & \begin{sideways}\textit{Negative}\end{sideways} \\
\arrayrulecolor{black}\hline
\textit{E1} &
&&&&
&&
& \checkmark &&&
\checkmark &&\\
\arrayrulecolor{lightgray}\hline
\textit{E2} &
&&&&
&&
& \checkmark &&&
\checkmark &&\\
\arrayrulecolor{lightgray}\hline
\textit{E3} &
&&&&
&&
&& \checkmark &&
& \checkmark &\\
\arrayrulecolor{lightgray}\hline
\textit{E4} &
&&&&
&&
&&& \checkmark &
&& \checkmark \\
\arrayrulecolor{lightgray}\hline
\textit{E5} &
&&&&
&&
& \checkmark &&&
\checkmark &&\\
\arrayrulecolor{lightgray}\hline
\textit{E6} &
\checkmark &&&&
\checkmark &&
&& \checkmark &&
\checkmark &&\\
\arrayrulecolor{lightgray}\hline
\textit{E7} &
& \checkmark &&&
\checkmark &&
&& \checkmark &&
&& \checkmark \\
\arrayrulecolor{lightgray}\hline
\textit{E8} &
&&&&
&&
&& \checkmark &&
& \checkmark &\\
\arrayrulecolor{lightgray}\hline
\textit{E9} &
\checkmark &&&&
& \checkmark &
& \checkmark &&&
\checkmark &&\\
\arrayrulecolor{lightgray}\hline
\textit{E10} &
\checkmark &&&&
& \checkmark &
&& \checkmark &&
\checkmark &&\\
\arrayrulecolor{lightgray}\hline
\textit{E11} &
& \checkmark &&&
\checkmark &&
& \checkmark &&&
\checkmark &&\\
\arrayrulecolor{lightgray}\hline
\textit{E12} &
\checkmark &&&&
\checkmark &&
\checkmark &&&&
\checkmark &&\\
\arrayrulecolor{lightgray}\hline
\textit{E13} &
&&&&
&&
\checkmark &&&&
\checkmark &&\\
\arrayrulecolor{lightgray}\hline
\textit{E14} &
& \checkmark &&&
\checkmark &&
& \checkmark &&&
\checkmark &&\\
\arrayrulecolor{lightgray}\hline
\textit{E15} &
&&&&
&&
& \checkmark &&&
\checkmark &&\\
\arrayrulecolor{lightgray}\hline
\textit{E16} &
\checkmark && \checkmark &&
& \checkmark &
&& \checkmark &&
&& \checkmark \\
\arrayrulecolor{lightgray}\hline
\textit{E17} &
&&& \checkmark &
\checkmark &&
&& \checkmark &&
& \checkmark &\\
\arrayrulecolor{lightgray}\hline
\textit{E18} &
&&&&
&&
& \checkmark &&&
\checkmark &&\\
\arrayrulecolor{lightgray}\hline
\textit{E19} &
&&&&
&&
& \checkmark &&&
& \checkmark &\\
\arrayrulecolor{lightgray}\hline
\textit{E20} &
&&&&
&&
\checkmark &&&&
\checkmark &&\\
\arrayrulecolor{lightgray}\hline
\textit{E21} &
\checkmark &&&&
& \checkmark &
& \checkmark &&&
\checkmark &&\\
\arrayrulecolor{lightgray}\hline
\textit{E22} &
&&&&
&&
&& \checkmark &&
& \checkmark &\\
\arrayrulecolor{lightgray}\hline
\textit{E23} &
&&&&
&&
&& \checkmark &&
& \checkmark &\\
\arrayrulecolor{black}\hline
\textbf{Totals} & 6 & 3 & 1 & 1 & 6 & 4 & 3 & 10 & 9 & 1 & 14 & 6 & 3 \\
\arrayrulecolor{black}\hline
\end{tabular}
\caption{Demographic backgrounds and prior experiences with LLMs of the educators that we interviewed.}
\label{t-interviewees-educator-individual2}
\Description{A table describing the individual demographic backgrounds and prior experiences with LLMs of the educators we interviewed. E1 uses LLMs regularly and has a positive opinion of them. E2 uses LLMs regularly and has a positive opinion of them. E3 has limited prior use of LLMs and has a mixed opinion of them. E4 has never used LLMs and has a negative opinion of them. E5 uses LLMs regularly and has a positive opinion of them. E6 is a white man who has limited prior use of LLMs and has a positive opinion of them. E7 is a Black man who has limited prior use of LLMs and has a negative opinion of them. E8 has limited prior use of LLMs and has a mixed opinion of them. E9 is a white woman who uses LLMs regularly and has a positive opinion of them. E10 is a white woman who has limited prior use of LLMs and has a positive opinion of them. E11 is a Black man who uses LLMs regularly and has a positive opinion of them. E12 is a white man who has significant prior use of LLMs and has a positive opinion of them. E13 has significant prior use of LLMs and has a positive opinion of them. E14 is a Black man who uses LLMs regularly and has a positive opinion of them. E15 uses LLMs regularly and has a positive opinion of them. E16 is a white and Middle Eastern woman who has limited prior use of LLMs and has a negative opinion of them. E17 is a Hispanic man who has limited prior use of LLMs and has a mixed opinion of them. E18 uses LLMs regularly and has a positive opinion of them. E19 uses LLMs regularly and has a mixed opinion of them. E20 has significant prior use of LLMs and has a mixed opinion of them. E21 is a white woman who uses LLMs regularly and has a mixed opinion of them. E22 has limited prior use of LLMs and has a mixed opinion of them. E23 has limited prior use of LLMs and has a mixed opinion of them.}
\end{table*}

\end{document}